# Anomalous lattice anharmonicity and spin-lattice coupling in spin-orbit coupled halide $K_2IrBr_6$

Smriti Bhatia,[a] Arshid Ahmad [a,b], Mohd Zeeshan,[a] Simranjeet Kaur,[a] Reshma Kumawat[a], Shubham Farswan[a], Jay Krishna Anand,[c], Ankur Goswami,[c] Brajesh K Mani,[a] Kaushik Sen,*[a]

The interplay between lattice distortions, magnetism, and spin-orbit coupling (SOC) in 5d transition-metal halides offers a fertile platform for exploring correlated spin-lattice dynamics. Here, we investigate the impact of structural symmetry breaking on lattice vibrations and local spin environments in the antifluorite compound $K_2IrBr_6$ using temperature-dependent Raman spectroscopy, electron paramagnetic resonance (EPR), and first-principles lattice dynamics calculations. $K_2IrBr_6$ undergoes successive cubic-to-tetragonal and tetragonal-to-monoclinic phase transitions at ~170 K and ~122 K, respectively, driven by cooperative distortions of the $IrBr_6$ octahedra. Raman spectroscopy reveals anomalous phonon linewidth broadening and unconventional temperature dependence of phonon energies near these transitions, indicating that dynamic spin-phonon coupling is significant well above the Néel temperature ($T_N \approx 16$ K). First-principles phonon calculations support the mode assignments and demonstrate that symmetry-lowering distortions significantly renormalize vibrational modes, consistent with the experimental observations. Complementary EPR measurements detect anisotropic $g$-factors, resonance field shifts, and linewidth narrowing across the structural transitions, reflecting the emergence of static spin-lattice correlations mediated by spin-orbit entanglement. These findings establish $K_2IrBr_6$ as a model system where halide ligand fields, octahedral distortions, and SOC collaboratively govern spin-lattice coupling, providing chemical pathways to engineer quantum materials with tunable magnetic and lattice responses.

[a.] Department of Physics, Indian Institute of Technology Delhi, Hauz Khas, New Delhi 110016, India
[b.] Institute of Condensed Matter Physics, Technische Universität Darmstadt, 64289 Darmstadt, Germany
[c.] Department of Materials Science and Engineering, Indian Institute of Technology Delhi, Hauz Khas, New Delhi-110016, India
* kaushik.sen@physics.iitd.ac.in

Footnotes relating to the title and/or authors should appear here.

## Introduction

The interplay between lattice distortions, magnetism, and spin-orbit coupling (SOC) in $5d$ transition metal compounds has garnered significant attention due to its role in stabilizing exotic quantum phases.[1,2] Unlike their $3d$ counterparts, $5d$ materials exhibit a delicate competition between SOC, Coulomb interactions, and crystal field effects, often leading to unconventional magnetic and lattice dynamics.[3,4,5] Understanding how structural distortions influence both spin and lattice degrees of freedom is crucial for unravelling emergent behaviors in SOC-driven systems, where spin-lattice interactions and phonon anomalies become intertwined.[1,6]

Materials such as $Sr_2IrO_4$ and $Na_2IrO_3$ have demonstrated that strong SOC can stabilize $j_{\mathrm{eff}} = 1/2$ electronic states, resulting in novel magnetic ground states.[7,8] More recently, perovskite and antifluorite-type iridates have emerged as promising platforms to explore spin-lattice coupling, with examples such as $Ba_2YIrO_6$ and $K_2IrCl_6$ revealing intricate connections between structural distortions and magnetic or vibrational properties.[9,10] Among these, the antifluorite family $K_2IrX_6$ (X = Cl, Br, I) is particularly intriguing because its face-centered cubic (FCC) lattice provides a nearly ideal realization of $j_{\mathrm{eff}} = 1/2$ magnetism. In these materials, octahedral tilts, rotations, and cooperative deformations significantly impact phonon behavior, exchange interactions, and magnetic anisotropy. $K_2IrCl_6$ has recently been proposed as a candidate for a nodal-line spin-liquid phase, where SOC and exchange frustration stabilize extended spin correlations without long-range magnetic order.[11]

$K_2IrBr_6$ presents an excellent opportunity to investigate the coupling between spin, lattice, and structural symmetry changes. It undergoes successive structural transitions from a cubic phase to a tetragonal phase at 170 K and further to a monoclinic phase at 122 K, driven by distortions of the $IrBr_6$ octahedra.[12] These structural changes accompany a transition from a high-temperature paramagnetic phase to an antiferromagnetic ground state below $T_N \approx$ 12 K, suggesting a strong coupling between lattice symmetry breaking and spin dynamics.

While prior studies have established the structural phase diagram and magnetic ordering in $K_2IrBr_6$, the microscopic interplay between lattice distortions and spin dynamics remains largely unexplored. In particular, it is unclear how symmetry-lowering transitions influence both lattice vibrations and the local magnetic environment, or whether evolving spin correlations reciprocally affect lattice properties. Furthermore, whether these spin-lattice couplings persist in the paramagnetic regime, well above $T_N$, remains an open question.

In this work, we investigate the spin-lattice coupling in $K_2IrBr_6$ by combining temperature-dependent Raman spectroscopy and electron paramagnetic resonance (EPR) spectroscopy. Raman spectroscopy reveals pronounced anomalies in phonon linewidths and energies across the structural transitions, indicating that spin-phonon coupling is significant well above $T_N$. These dynamic couplings manifest as enhanced lattice anharmonicity, reflecting the interplay between phonons and short-range antiferromagnetic (AF) fluctuations. Complementary EPR measurements probe the static magnetic environment of $j_{\mathrm{eff}} = 1/2$ spins, revealing anisotropic $g$-factors and linewidth narrowing in the low-symmetry phases, consistent with the persistence of spin fluctuations and measurable spin–phonon coupling effects above $T_N$. Together, these findings establish $K_2IrBr_6$ as a model system where structural distortions, SOC, and spin correlations are deeply interlinked, providing insight into complex lattice dynamics and local magnetism in halide-based SOC-driven materials.

## Experimental methods and computation

### Sample and structural characterization

We performed experiments on commercially available polycrystalline $K_2IrBr_6$ (KIB) (Ir 25.4% min) from Alfa Aesar. We performed symmetric $\boldsymbol{\theta}$-$\boldsymbol{2\theta}$ powder x-ray diffraction (XRD) experiments at room temperature using a Malvern Panalytical Empyrean Series 3 diffractometer with Cu-Kα radiation ($\boldsymbol{\lambda} = \mathbf{1.54}$ Å). Fig. 1(a) shows the XRD data along with a Rietveld refinement performed considering the reported cubic phase ($\boldsymbol{Fm\bar{3}m}$) near room temperature.[12] The fair agreement with the data testifies to the high crystalline quality and phase purity of the investigated polycrystalline sample. Previously, high-resolution XRD experiments using synchrotron radiation ($\boldsymbol{\lambda}$ =0.4001 Å) were performed to investigate structural phase transitions in the compound and the corresponding lattice parameters in the cubic, tetragonal, and monoclinic phases were reported.[12] From our analysis, we obtained a = b = c = 10.28 Å at room temperature, compared to the reported 10.29 Å at 270 K. Fig. 1(b) shows the rietveld refined cubic structure of KIB, specifically adopting a face-centered cubic arrangement with the $Ir^{4+}$ ions at the vertices and faces of the cubic unit cell.

### Magnetization measurement

DC magnetization measurements were carried out using a Quantum Design made MPMS-3 setup. For measurements, we wrapped 15 mg polycrystalline sample in a Teflon tape and placed it inside a straw that is attached to the sample rod via an adapter. We punctured the straw in several places near the sample to facilitate purging inside the helium gas environment to reduce the trapped air inside the sample, which often gives rise to an unwanted signal around 50 K. For

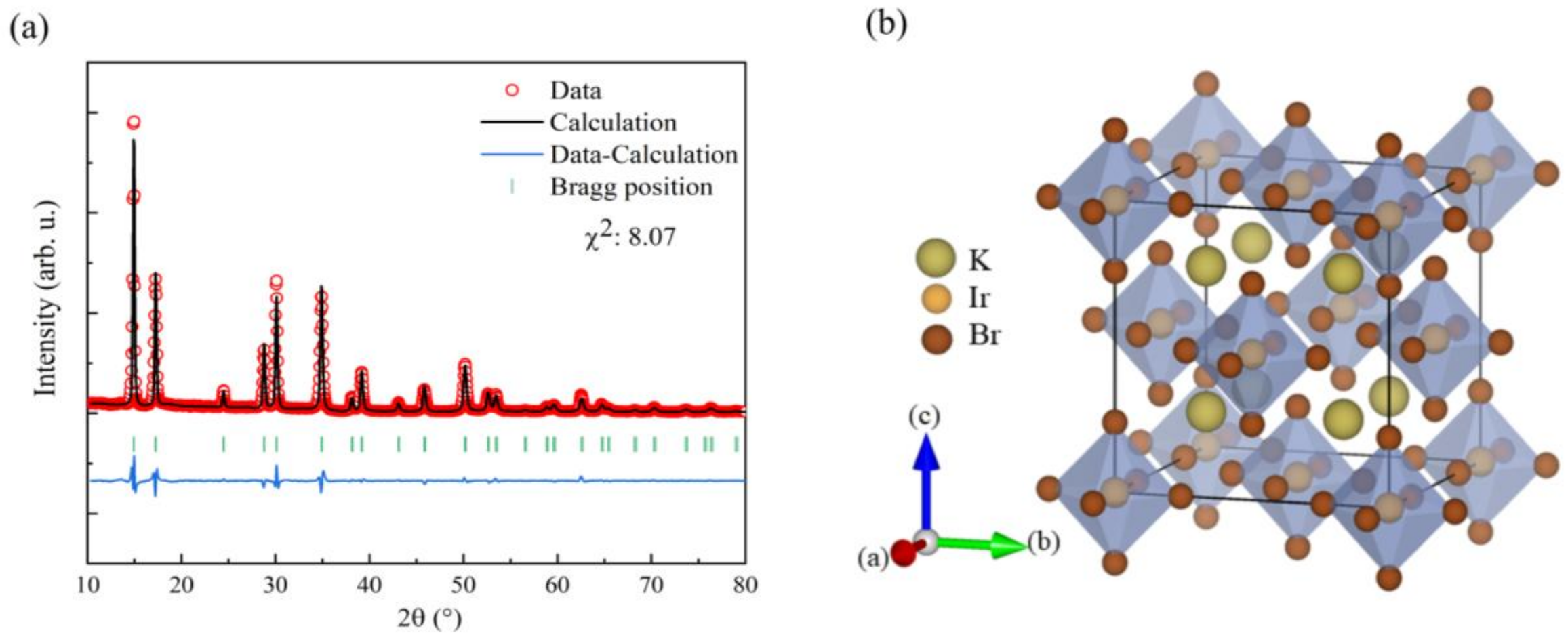

**Fig. 1** (a) Symmetric $\theta - 2\theta$ x-ray diffraction scan at room temperature, along with Rietveld refinement, confirming the cubic structural phase. (b) Schematic diagram of cubic unit cell of K2IrBr6 for Rietveld refined structure at room temperature.

data acquisition, we cooled from 300 to 2 K under an applied magnetic field of 0.1 tesla and recorded the data while warming up to 300K.

#### Low-temperature Raman scattering

We performed unpolarized Raman spectroscopy experiments using a home-built setup based on a Horiba-made iHR550 grating spectrometer. For laser excitation, a He-Ne laser (632.8 nm) was focused through a $10\times$ long-working-distance objective onto the sample. A spectral resolution of 2.5 $cm^{-1}$ was achieved using a grating with 1200 grooves/mm. Two notch filters were used to suppress the Rayleigh scattering, resulting in the lowest reliable energy cut-off of 80 $cm^{-1}$.

For low-temperature measurements between 11 and 300 K, we used an ultralow-vibration microscopy cryostat from ColdEdge Technologies. The laser power was kept below 1 mW to avoid laser-induced heating at low temperatures. The obtained Raman spectra consist of a broad background and a set of sharp peaks arising from lattice vibrations (phonons). To analyze these phonon modes, specifically to determine their Raman shift and linewidth, we fit the data using a sloping Gaussian background and Voigt profiles for the phonon peaks. In the Voigt profile, the Gaussian width was fixed to the instrumental resolution of our Raman setup, while the Lorentzian linewidth was treated as a free parameter to deconvolute the intrinsic phonon linewidth from the instrument resolution.

#### Low-temperature electron paramagnetic resonance

Electron paramagnetic Resonance (EPR) spectra were recorded as a function of temperature using Bruker Biospin spectrometer (A300-9.5/12/S/W). The powder was kept in the quartz sample tube, which was placed in the EPR cavity for measurement. The data were recorded at the microwave frequency of 9.39 GHz at a power of 1 mW. The data were accumulated seven times at each temperature for better data quality. EPR results were recorded at several temperatures while warming up from 100 K to 300 K.

#### First-principles lattice dynamics calculations

The structural, vibrational, and symmetry properties of KIB were investigated using density functional theory (DFT) and density functional perturbation theory (DFPT) as implemented in the Vienna Ab initio Simulation Package (VASP),[13,14] within the projector augmented-wave (PAW) method.[15] The Perdew-Burke-Ernzerhof (PBE) generalized gradient approximation (GGA) was used for the exchange–correlation functional.[16] A plane-wave cutoff energy of 500 eV was employed, the energy convergence was set at $10^{-6}$ eV, and the forces on all atoms were minimized to converge at $10^{-4}$ eV/Å. The initial structural relaxations were carried out for the cubic, tetragonal, and monoclinic phases of KIB using a Monkhorst-Pack $k$-grid of 11×11×11 for cubic and 11x11x9 for rest of the phases. A denser $k$-grid of 21x21x21 and 15x15x11 were used in respective order for self-consistent energy calculations. Spin-orbit coupling (SOC) was incorporated to account for the relativistic effects. The strongly correlated $d$-electrons of Ir were accounted using DFT+U approach (Dudarev scheme).[17] The Hubbard U value of 1.8 eV was applied on Ir, consistent with literature.[12] The obtained electronic structure showed bandgap of 0.30, 0.27, and 0.31 eV, respectively for the cubic, tetragonal, and monoclinic space groups. Symmetry-preserving supercells were generated to accommodate finite-displacement calculations. For the cubic phase, a 2×2×1 supercell containing 144 atoms was constructed. Similarly, 2×2×1 supercells with 72 atoms each were used for the tetragonal and monoclinic phases. The dynamical matrix was computed using finite-displacement methods. This involved small displacements of atoms 0.01 Å within the supercell, and the resulting forces were used to calculate phonon frequencies. A $\Gamma$- centered $k$ - point mesh of 4×4×4 was used for these calculations to ensure accurate results for subsequent simulations. The Phonopy package was employed to determine phonon frequencies and displacement vectors.[18] These calculations were performed for high-symmetry points in the Brillouin zone to examine the phonon dispersion and mode classification. A detailed symmetry analysis was conducted to identify Raman- and IR-active phonon modes for each phase. This step provided insight into the vibrational properties and their potential experimental observability.

## Results and Discussions

### Magnetic properties

Fig. 2 shows DC magnetic susceptibility ($\chi$) of polycrystalline KIB under a magnetic field of 0.1 T. $\chi$ increases with decreasing temperature down to 16 K, indicating a paramagnetic state above 16 K. The continuous drop in $\chi$ below ~16 K towards low temperatures is an indication of the occurrence of antiferromagnetic (AF) order with $T_N \cong 16$ K which is close to the reported value of 12 K.[12] In the inset, we plotted the temperature derivative of the inverse susceptibility $(d\chi^{-1}/dT)$ between 150 and 220 K. The distinct peak around 184 K in $(d\chi^{-1}/dT)$ versus $T$ plot has been associated with the first-order cubic-to-tetragonal phase transition. This is consistent with the previous report.[12]

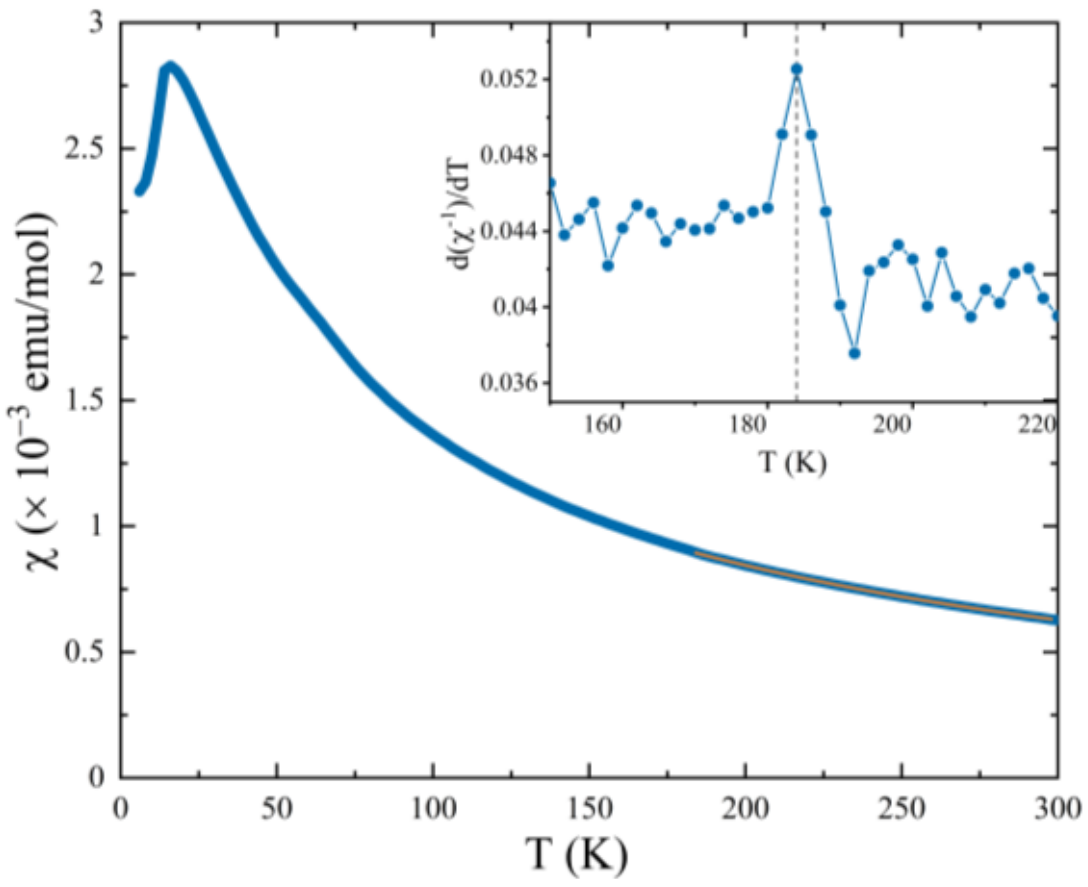

**Fig. 2** DC magnetic susceptibility ($\chi$) versus temperature for polycrystalline $K_2IrBr_6$ ($\mu_0H$=0.1 T). The inset shows the d($\chi^{-1}$)/dT in the vicinity of cubic to tetragonal transition.

$\chi$ vs. $T$ well above the cubic-to-tetragonal structural transition (184 K) in the paramagnetic phase was fitted with the following Curie-Weiss formula as mentioned in equation 1.

$$\chi = \chi_0 + \frac{C}{T - \theta_{CW}} \quad (1)$$

where $\chi_0$ is the temperature-independent diamagnetic contribution, $C$ is the Curie constant and $\theta_{CW}$ is the Curie-Weiss temperature. Here, $C = \frac{N_A \mu_{eff}^2}{3k_B}$ , where $N_A$ is Avogadro's number, $\mu_{\text{eff}}$ is the effective paramagnetic moment, and $k_B$ is the Boltzmann constant. The fitting yields $\chi_0 = -9.6 \times 10^{-4}$ emu/mol, which can be compared with $\chi_0$ for other $Ir^{4+}$ compounds.[10,19] The fitted value of Curie constant, $C$ = 0.2438 emu K/mol, and $\theta_{CW}$ = -88.6 K, which are in close agreement with the reported values by Khan et al.[12] Further, we obtained $\mu_{\text{eff}} = 1.38$ μB per $Ir^{4+}$ ion, which is close to the expected magnetic moment of 1.73 μB for the total angular momentum of $j_{\text{eff}} = 1/2$ and the Landé $g$-factor of $g = 2$.

### Calculated phonon modes

To investigate how lattice dynamics evolve across structural transitions, we first establish the expected Raman phonon modes at the zone center through first-principles calculations, as no prior Raman studies exist on KIB. Group theory analysis for the cubic phase (space group Fm$\bar{3}$m) yields one $A_{1g}$, one $E_g$, and two $T_{2g}$ phonon modes under the point group of $O_h$. Reduction of lattice symmetry from cubic to tetragonal (P4/mnc) and further to monoclinic (P21/n) gives rise to increased numbers of phonon modes. The tetragonal phase yields three $A_{1g}$, three $B_{1g}$, two $B_{2g}$, and six $A_g$ modes under the point group of $D_{4h}$. In the monoclinic phase, the phonon modes are twelve Ag and twelve $B_g$ under the point group of $C_{2h}$. These are summarized in Table I for brevity.

We determined the energy of the phonon modes and their displacements using first-principles lattice dynamics calculations based on density functional perturbation theory (DFPT). Table II lists the phonon mode energies for the cubic phase (phonon mode energies for the tetragonal and monoclinic phases are given in supplementary information for completeness). The corresponding phonon displacement vectors are shown in Fig. 3(a)-(d). The $T_{2g}$ modes in cubic KIB include two distinct displacement patterns. The external $T_{2g}$ mode (61.6 $cm^{-1}$) represents the relative motion of K ions with respect to the nearly stationary $IrBr_6$ octahedra. This motion resembles a lattice-breathing effect, with the K-atoms oscillating within the framework defined by the rigid octahedra. The internal $T_{2g}$ mode (108.3 $cm^{-1}$) involves the scissoring motion of the Br-atoms within the $IrBr_6$ octahedra. These vibrations tangentially displace the Br-atoms, causing a twisting or shearing motion of the octahedra without significantly altering the bond lengths between the Ir- and Br-atoms. The $E_g$ mode (189.8 $cm^{-1}$) involves symmetric stretching of opposite Br-atoms towards or away from the central Ir atom. This motion introduces transient distortions in the $IrBr_6$ octahedron, leading to dynamic tetrahedral strains without strictly preserving the ideal octahedral symmetry. The $A_{1g}$ mode (222.4 $cm^{-1}$) corresponds to the symmetric stretching of the Br atoms in the $IrBr_6$ octahedra, with all Br atoms moving radially, in phase, towards or away from the central Ir atom, preserving the octahedral symmetry.

Table I. Raman-active optical phonon mode symmetries at zone-center for cubic, tetragonal, and monoclinic phases.

| Structure<br>Point Group | Cubic<br>$O_h$ | Tetragonal<br>$D_{4h}$ | Monoclinic<br>$C_{2h}$ |
|---|---|---|---|
| $\Gamma_{Raman}$ | $A_{1g}+E_g+2T_{2g}$ | $3A_{1g}+3B_{1g}+2B_{2g}+6E_g$ | $12A_g+12B_g$ |

Table II. Comparison of the calculated and experimental frequencies for Raman phonon modes in the cubic phase of $K_2IrBr_6$ (300K).

| $\Gamma_{Raman}$ | $\omega_{cal}$ (cm$^{-1}$) | $\omega_{exp}$ (cm$^{-1}$) | $\Delta\omega = \frac{\lvert\omega_{cal}-\omega_{exp}\rvert}{\omega_{cal}} \times 100$ (%) |
|---|---|---|---|
| $T_{2g}$ | 61.6 | - | - |
| $T_{2g}$ | 108.3 | 111.3 | 2.78 |
| $E_g$ | 189.8 | 174.7 | 7.95 |
| $A_{1g}$ | 222.4 | 212.3 | 4.54 |

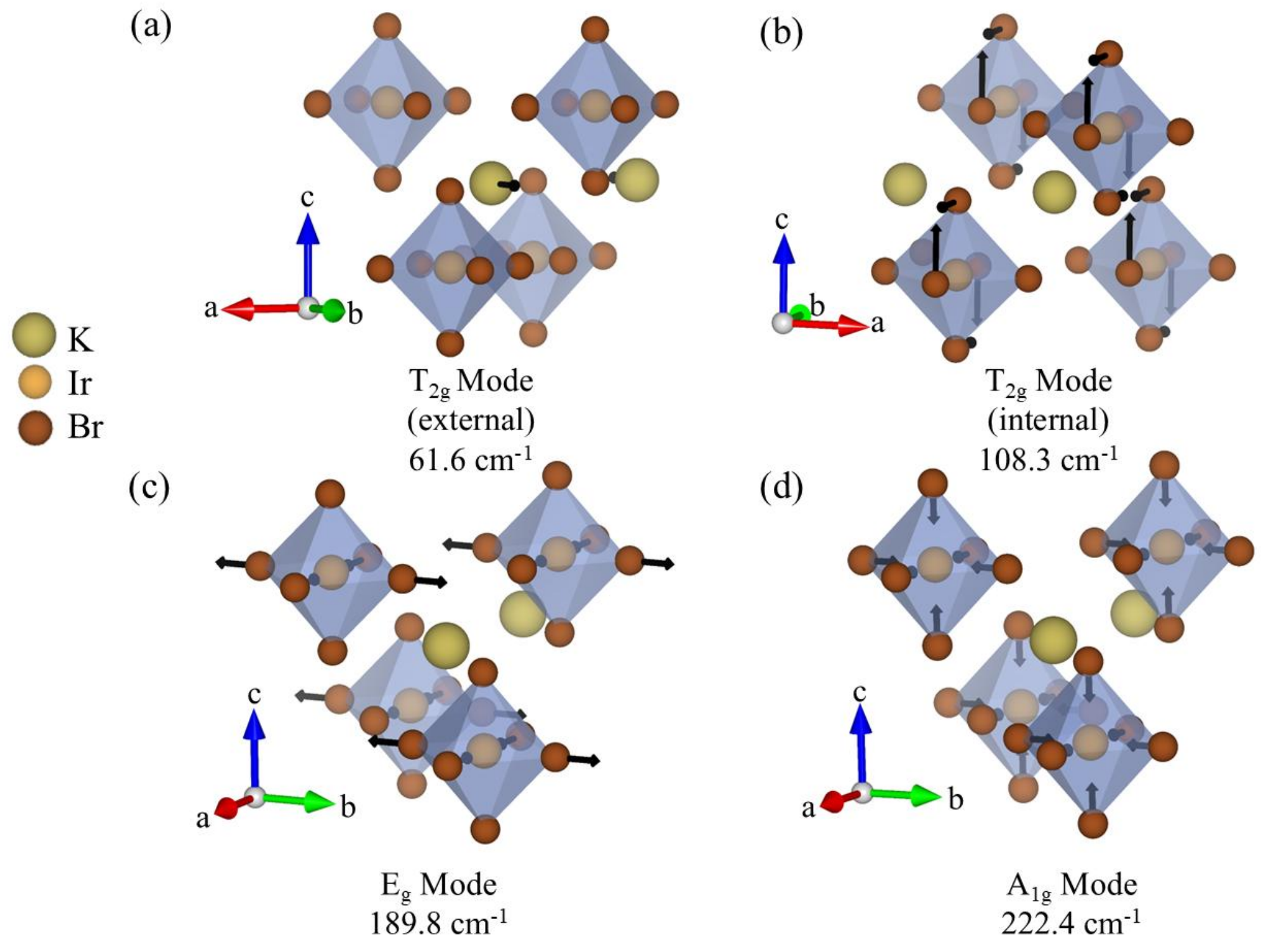

**Fig. 3** Phonon displacement vectors for the four Raman-active phonon modes in cubic symmetry of KIB. (a)$T_{2g}$ (external), (b) $T_{2g}$ (internal), (c) $E_g$, and (d) $A_{1g}$.

**Raman scattering study of phonon modes**

Fig. 4 shows unpolarized Raman spectra of KIB at several temperatures starting from 11 to 300 K. The spectrum at room temperature shows three prominent modes at 111.3, 174.7, and 212.3 $cm^{-1}$. From unpolarized Raman scattering experiments alone, we cannot assign the mode symmetries. However, the proximity of these modes with the calculated phonon mode energies, as listed in Table II, allows us to assign the mode symmetry. Specifically, the modes at 111.3, 174.7, and 212.3 $cm^{-1}$ are $T_{2g}$ (internal), $E_g$, and $A_{1g}$, respectively. The deviation of the calculated phonon mode energy from the experimental values is nominal, below 8 %. This testifies to the reasonably high accuracy of our calculations, which facilitates the mode assignment. Note that we were not able to detect the expected fourth mode, i.e. $T_{2g}$ (external) at 61.6 $cm^{-1}$, as the lowest energy cut-off in our Raman setup is limited to $\sim$ 80 $cm^{-1}$.

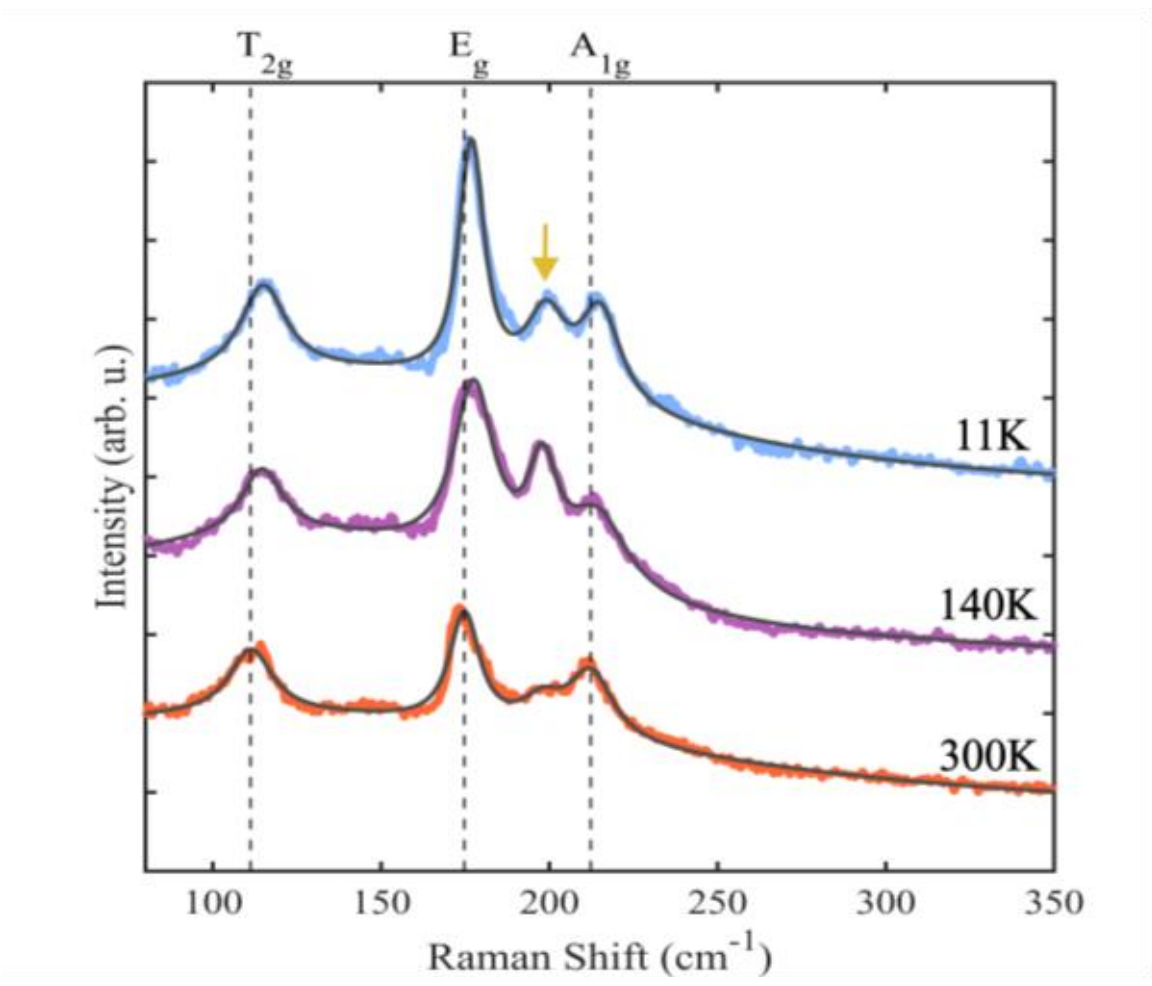

**Fig. 4** Representative unpolarized Raman spectra of KIB at selected temperatures. The dark solid lines indicate the Voigt fits at the respective temperatures.

We now focus on the low-temperature Raman spectra, particularly in the tetragonal and monoclinic phases, as shown for 140 K (tetragonal) and 11 K (monoclinic) in Fig. 4. A mode near 200 $cm^{-1}$ emerges as the system enters the tetragonal phase and persists into the monoclinic phase. Based on group theoretical analysis (Table I) and first-principles phonon calculations (see Supplementary Information), additional Raman-active modes are expected at low temperatures due to reduced crystallographic symmetry. In the tetragonal phase, three modes are predicted near 200 $cm^{-1}$: 189.1 $cm^{-1}$ ($B_{1g}$), 189.5 $cm^{-1}$ ($B_{2g}$), and 190.8 $cm^{-1}$ ($A_{1g}$), differing from the observed mode by approximately 5-6%. However, given the instrumental resolution of 2.5 $cm^{-1}$, we are unable to resolve these modes individually and cannot determine whether a single mode or a superposition of modes is observed at 200 $cm^{-1}$.

In the monoclinic phase, two modes are expected in this spectral region: 194.0 $cm^{-1}$ ($A_g$) and 198.4 $cm^{-1}$ ($B_g$), yet no additional features beyond the four modes seen in the tetragonal phase are observed. Several factors may explain the absence of further resolved modes despite the lower symmetry. While first-principles calculations predict more Raman-active modes in the low-temperature phases, many may have negligible contributions to the polarizability tensor, rendering them too weak to detect. This is consistent with previous observations in isostructural compounds such as $K_2IrCl_6$ and $K_2ReCl_6$,[20,21] where symmetry-lowered modes appear with very low intensity. Furthermore, Raman scattering primarily probes collective lattice vibrations and thus reflects the global crystallographic symmetry. Local structural distortions may not be resolved if their contribution averages out over the probing volume.[21,22] Consequently, intense parent cubic-phase modes may overshadow weaker features, and phonon broadening due to anharmonicity may further obscure low-intensity signals.

To explore the microscopic lattice dynamics across structural phase transitions in KIB, we examined the temperature dependence of phonon energies ($\omega$) and linewidths (full width at half maximum) using Raman spectroscopy. Phonons, being sensitive to the structural and electronic environment of the material, can in some cases couple to other degrees of freedom such as spin or charge fluctuations. Therefore, tracking their temperature evolution provides a valuable route to investigate not only anharmonic phonon-phonon interactions but also possible couplings to correlated excitations. Fig. 5(a)-(h) shows the temperature evolution of Raman-active phonon modes across the cubic-to-tetragonal-to-monoclinic structural transitions. The data reveal distinct deviations from conventional anharmonic behaviour, suggesting the presence of additional scattering channels beyond phonon-phonon decay, and indicating that the nature

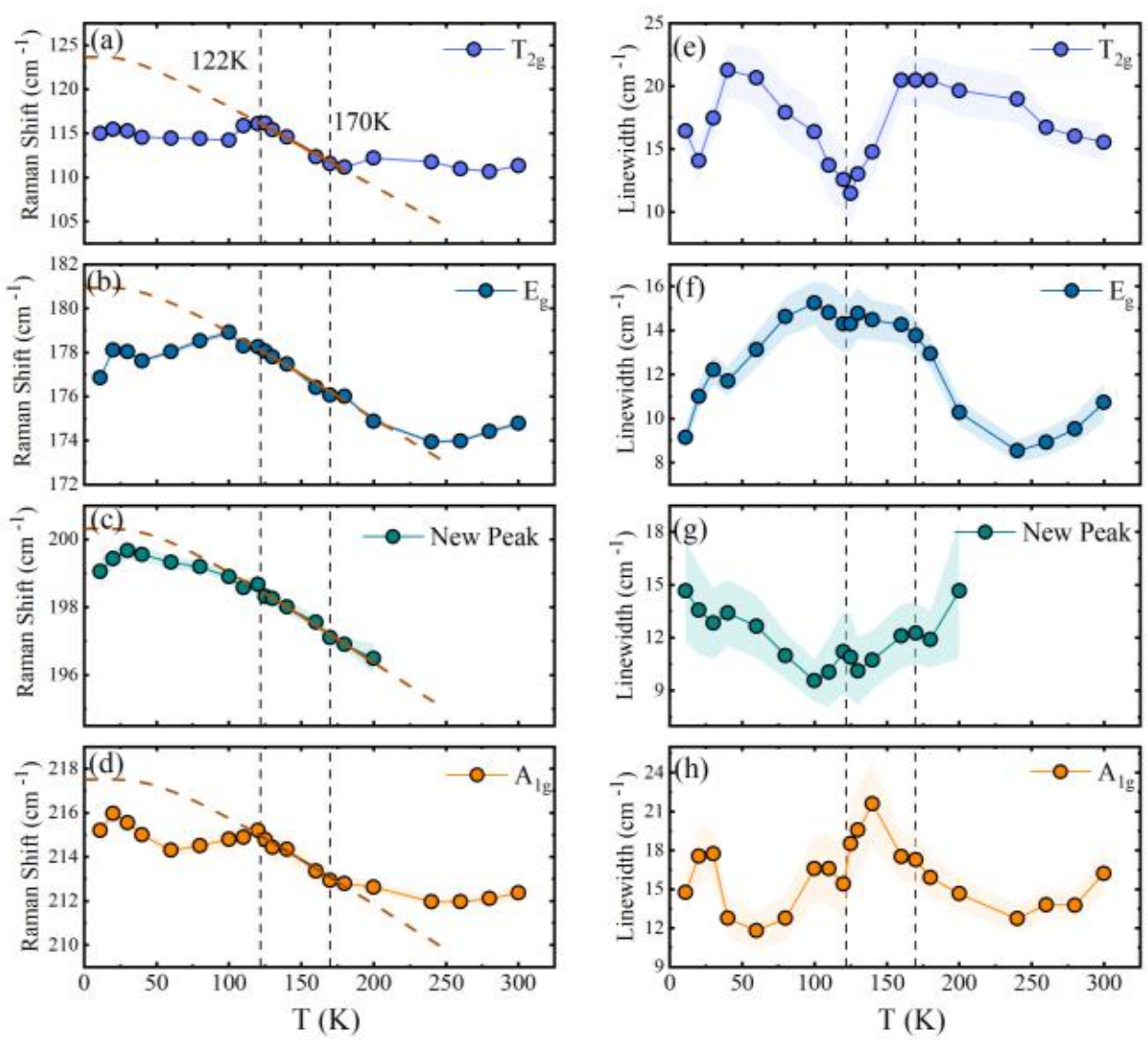

**Fig. 5** (a)-(d) Raman shift as a function of temperature (symbols) for $T_{2g}$ (internal), $E_g$, the new mode and $A_{1g}$ modes. (e)-(f) the corresponding linewidths as a function of temperature. The dashed brown line through the data in (a)-(d) are the best fits according to the Klemens decay model in the tetragonal phase. The vertical dotted lines are drawn corresponding to the phase transition temperatures as reported by Khan et al.[12] The color shade indicates the uncertainty range, i.e. the error bars. They were obtained from the nonlinear least-squares Voigt fitting routine.

and strength of anharmonic interactions evolve with changes in lattice symmetry.

All three original Raman-active modes: $T_{2g}$, $E_g$, and $A_{1g}$ exhibit prominent non-monotonic changes in both energy $(\omega)$ and linewidth $(\Delta\omega)$ as a function of temperature $(T)$. For the new mode emerging at low temperatures, $\omega$ vs. $T$ appears to be nearly monotonic, as shown in Fig. 5(c). For regular phonons, the temperature dependence of energy and linewidth is typically governed by anharmonic phonon decay, well described by the Klemens model.[23–26] In this framework, an optical phonon decays into two acoustic phonons of equal and opposite momentum. The resulting phonon-phonon scattering broadens linewidths and causes phonon softening with increasing temperature, often in a nearly linear fashion at high temperatures due to enhanced phonon population.

However, the experimental behaviour observed in KIB deviates significantly from these expectations. In the high-temperature cubic phase, the linewidths of the $T_{2g}$ and $A_{1g}$ modes remain nearly temperature-independent, while the $E_g$ mode exhibits anomalous broadening as temperature decreases. This contradicts the usual trend, where phonon-phonon scattering becomes weaker at lower temperatures, leading to narrower linewidths. On further cooling, abrupt changes are observed in all three modes, particularly across the structural transitions at 170 K and 122 K. Such abrupt changes in linewidth across structural phase transitions have also been observed by Nguyen et al. in halide perovskites.[27]

For the $E_g$ mode, anomalous linewidth broadening begins upon cooling below 250 K, which is well above the cubic-to-tetragonal transition at 170 K and persists through the tetragonal-to-monoclinic transition at 122 K, before narrowing again at lower temperatures. This behaviour suggests that phonon decay is influenced by additional interactions, likely involving other degrees of freedom. Since KIB is an insulator, electron-phonon scattering is unlikely. Instead, spin-phonon interactions are a more plausible mechanism.

Indeed, KIB undergoes long-range antiferromagnetic (AF) ordering below the Néel temperature ($T_N$ = 16 K), and the large negative Curie-Weiss temperature ($\theta_{CW}$ = –88.6 K) obtained from magnetization data indicates strong AF correlations well above $T_N$.[28] These short-range AF fluctuations in the paramagnetic phase can couple to lattice vibrations, opening a secondary phonon decay channel that enhances linewidths even above the magnetic ordering temperature (shorter lifetime leads to larger linewidth). The cusp-like feature in the $E_g$ linewidth near $T_N$ in Fig. 5(f) likely results from the interplay of two competing effects: decreasing phonon-phonon scattering and increasing spin-phonon scattering as the temperature approaches $T_N$ from above.

This interpretation is further supported by the phonon self-energy framework, where the complex self-energy $\Sigma = \Sigma' + i\Sigma''$ comprises real and imaginary components. The imaginary part $\Sigma''$ corresponds to the phonon lifetime and contributes directly to the linewidth, while the real part $\Sigma'$ influences the phonon energy.[29,30] Therefore, deviations in both linewidth and energy from the expected anharmonic trend point to additional interactions. For all three modes, phonon energies deviate from the Klemens model across the full temperature range of 11-300 K. Surprisingly, a narrow temperature window between 170 K and 122 K exhibits an approximately linear phonon energy dependence on temperature. We verified that the anomalous temperature dependence of the phonon frequency and linewidth cannot be accounted for within the quasi-harmonic model, i.e. cannot be explained simply by the volume change of the system across different phases, as confirmed using the temperature-dependent XRD data reported by Khan et al.[12]

According to Klemens model, the phonon frequency $(\omega)$ as a function of temperature $(T)$ follows eq. 2.

$$\omega(T) = \omega_0 - C\left(1 + \frac{2}{e^{\hbar\omega_0/2k_BT} - 1}\right) \qquad (2)$$

where $\omega_0$ is the natural phonon frequency in the absence of any anharmonic lattice interactions, and $C$ is the anharmonic constant that signifies the strength of anharmonic interactions. We fitted the observed linear $\omega(T)$ regions using eq. (2). The fits show that deviations occur near both structural transitions, and that the extrapolated harmonic limit $\omega_0$ lies above the measured values across the entire range. This suggests that phonons continue to soften down to 11 K, contrary to expectations for a conventional crystalline solid, where atomic vibrations become harmonic at low temperatures.

Such persistent softening is indicative of anomalous lattice anharmonicity, which may arise either from residual structural distortions or from coupling to spin fluctuations. Previous structural refinements suggest that the lattice distortions stabilize below the monoclinic transition,[12] making spin–phonon interactions the more likely source of this behavior. Thus, AF fluctuations, present even above $T_N$, may play a central role in modifying lattice dynamics at low temperatures.

### Local magnetism with electron paramagnetic resonance

We performed electron paramagnetic resonance (EPR) measurements as a function of temperature to probe the impact of structural phase transitions on the local magnetic environment and spin dynamics above $T_N$. Since spin and lattice degrees of freedom can be interlinked, especially through spin-orbit coupling, EPR offers a complementary probe to our Raman study.

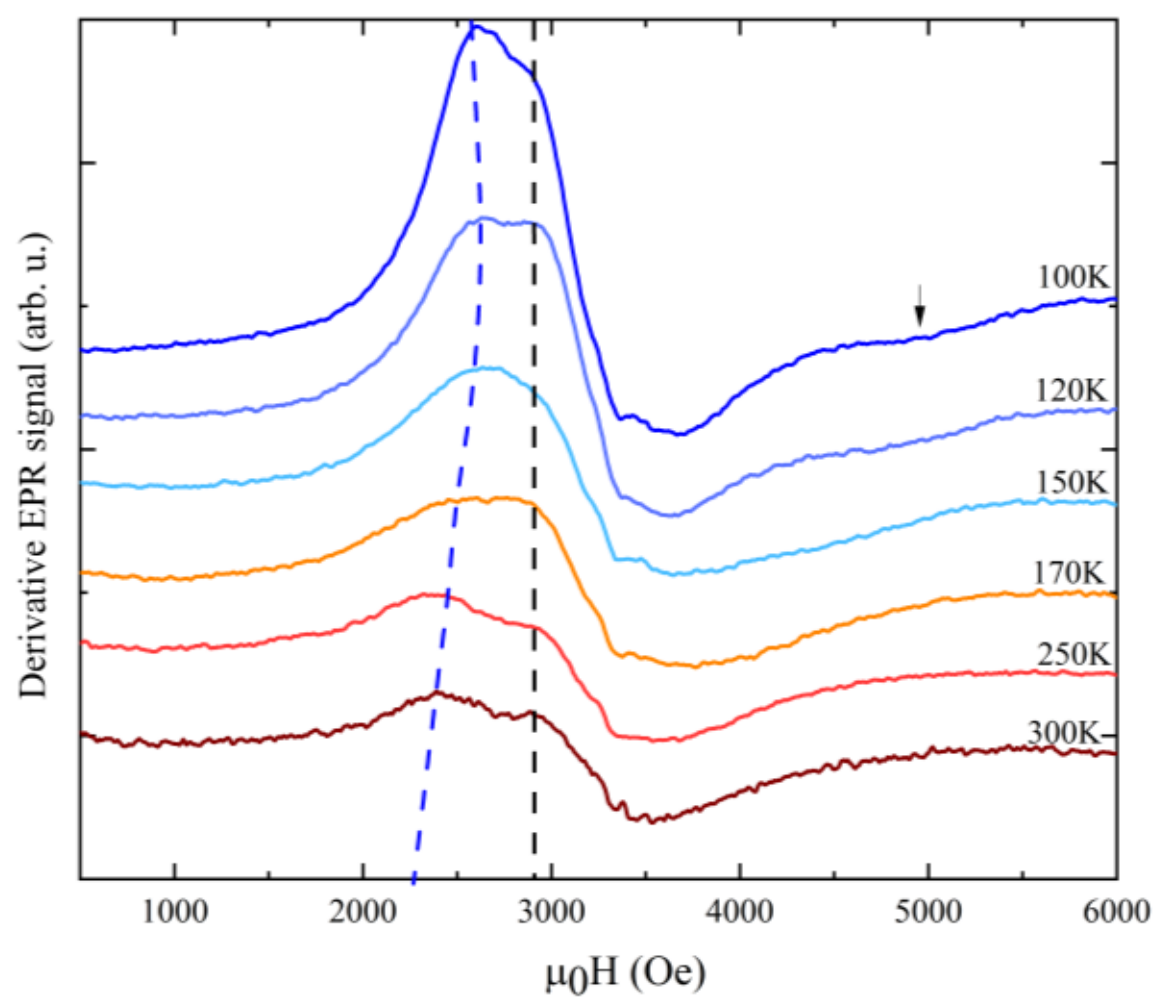

**Fig. 6** Derivative EPR spectra of polycrystalline KIB. The dashed lines are visual guides to highlight the substructures in EPR spectra.

Fig. 6 shows derivative EPR spectra ${dP}/_{dH}$ ($P$ is the absorbed microwave power and $H$ is the applied magnetic field) of polycrystalline KIB at several temperatures, from 100 to 300 K. At 300 K, in the cubic phase, two clear substructures are visible in the EPR spectrum, indicated by the dashed blue and black vertical lines. The substructure indicated by the blue dashed line is found to be the most sensitive to varying temperature. In particular, the mode shifts towards higher magnetic fields as we approach from the cubic to the tetragonal phase. Overall, the spectra get narrower towards low temperatures, which allows us to observe more substructures in the spectra. For instance, a curvature change appears around $\sim$5000 Oe at low temperatures, as indicated by an arrow in the 100 K spectrum. It seems that such a change in curvature begins to emerge already at 150 K (tetragonal phase). Certainly, the reduction in crystal symmetry from cubic to tetragonal, i.e. increased non-cubic distortions paves the way for the observation of this mode. Further, this high-field mode is significantly broader than the remaining substructures in the spectra.

The presence of multiple substructures in ${dP}/_{dH}$ confirms that $j_{\mathrm{eff}} = 1/2$ spins of $Ir^{4+}$ ions sit in a magnetically anisotropic environment. In contrast, for a magnetic isotropic system with uniform crystal-field effects, we expect a single EPR mode (single Lorentzian derivative profile), corresponding to a unique $g$-factor when polycrystalline sample is measured.[31] The pronounced anisotropic $g$-factor observed in our study reflects the strong spin-lattice coupling and substantial spin-orbit interaction inherent to the $Ir^{4+}$ ions. The robust spin-orbit coupling (SOC) in $5d$-$Ir^{4+}$ ensures that $j_{eff} = 1/2$ spins are not truly free but remain strongly entangled with the lattice through their orbital degrees of freedom. This inherent anisotropy is a hallmark of SOC-driven systems, where spin-orbit interactions induce significant deviations from ideal isotropy.[28,32,33]

After a qualitative discussion on the anisotropic $g$-factors, we now quantify them and track their temperature dependence and the lineshape of the corresponding EPR substructures. As mentioned earlier, the substructure indicated by the blue dashed line in Fig. 6 shows a relatively strong temperature dependence. To extract the lineshape parameters, i.e., resonant field ($\mu_0 H_r$), linewidth ($\mu_0 \Delta H$), and asymmetric spectral weight, the EPR spectra were fitted with the sum of the derivatives of symmetric ($L_S$) and anti-symmetric ($L_A$) Lorentzian profiles, along with a linear background term to account for minor baseline drifts.

The best fits were obtained when four modes are considered for the data at or below 150 K, while three modes are enough for the spectra above 150 K. We used the following functions for fitting:

$$I_{EPR} = \frac{1}{\mu_0}\left(\sum_{i=1}^{4} \frac{\partial}{\partial H}\left(L_{S,i} + L_{A,i}\right)\right) + mx + c \tag{2}$$

, where $\mu_0 H$ is the applied magnetic field, $mx + c$ is a linear background term to account for linear drift of the spectra and the summation runs for the four modes. $L_S$ and $L_A$ are defined as:

$$L_S = S\frac{\Delta H^2}{\Delta H^2 + (H - H_r)^2}, \tag{3}$$

$$L_A = A\frac{\Delta H^2 (H - H_r)}{\Delta H^2 + (H - H_r)^2} \tag{4}$$

, where $S$ and $A$ are the spectral weights for the symmetric and anti-symmetric Lorentzian profiles, respectively.

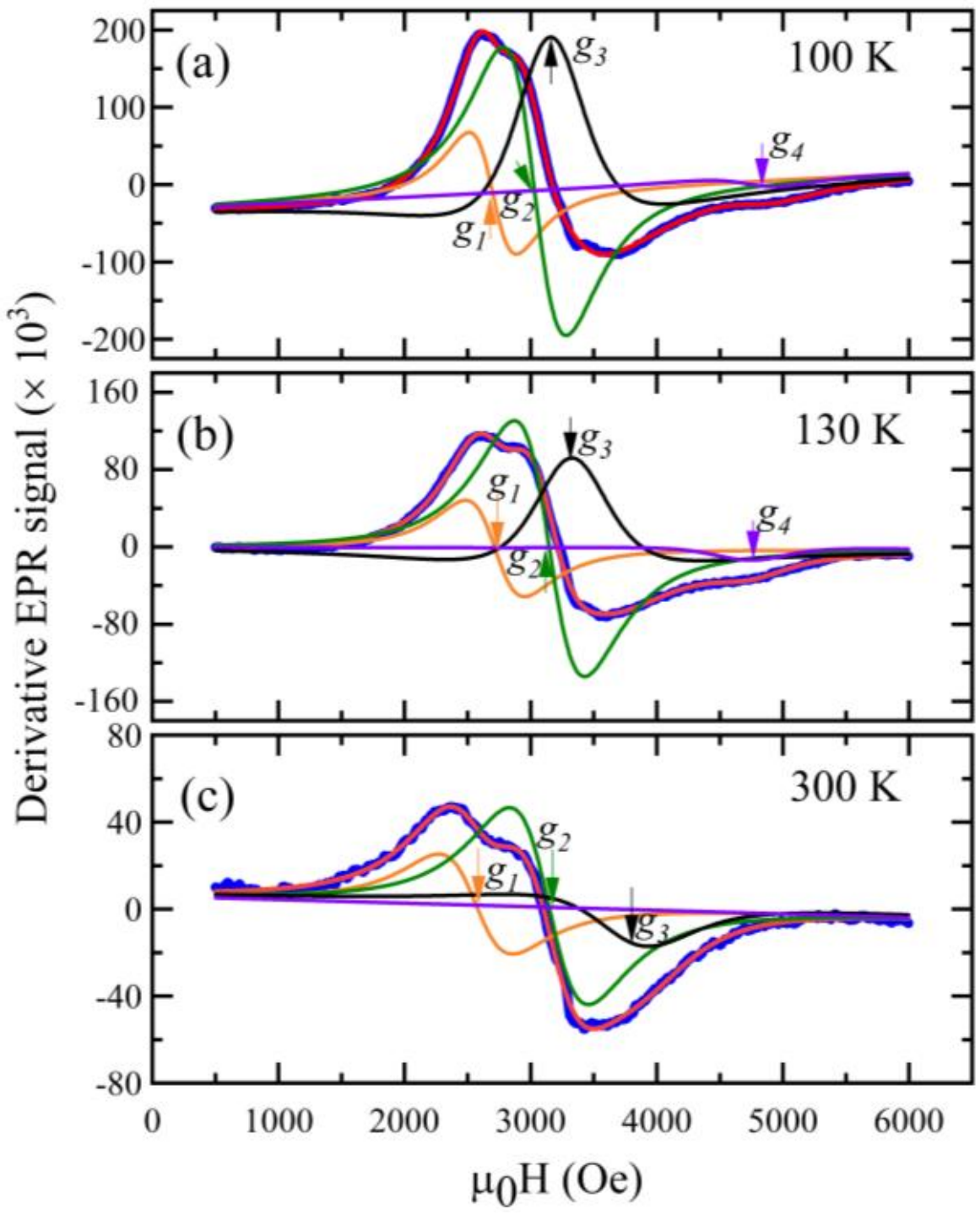

**Fig. 7** Representative fitting to the EPR spectra of polycrystalline KIB using eq. (2) at (a) 100, (b) 130, and (c) 300 K. Solid blue symbols are the raw data, and the solid red lines are the best fits. The substructure components are shown by solid lines, where orange: $g_1$, green: $g_2$, black: $g_3$, and violet: $g_4$.

Fig. 7(a), (b) and (c) show representative fits to the EPR spectra at three selected temperatures of 100, 130, and 300 K, respectively. As shown in Fig. 7, the EPR spectra, particularly the $g_3$-substructure (as indicated by arrow) gradually becomes more asymmetric and narrow with temperature reduction. This will be further discussed in the context of Fig. 9(b).

In accordance with the fitting results, we mark the resonant fields by arrows on the EPR spectra in Fig. 7. We mark the modes as $g_1$, $g_2$, $g_3$, and $g_4$ from low to high magnetic fields. Note that the significantly broad mode $g_4$ arises towards low temperatures. Although symmetry reduction appears to contribute, the precise origin of $g_4$ is not yet clear. The remaining $g$-factors correspond to $(g_x, g_y, g_z)$. Since we measure polycrystalline KIB, we cannot confirm $g$-values along crystallographic $x$, $y$, and $z$-axes.

Fig. 8 shows $\mu_0 H_r$ and $g$-factors for all the modes as a function of temperature. The $g$-factors were evaluated using the relationship $g = (h\upsilon/\mu_B\mu_0 H_r)$, where $h$ is the Planck's constant, $\mu_B$ is the Bohr magneton, and $\upsilon = 9.39$ GHz is the microwave frequency for $X$-band.[34] We marked cubic, tetragonal, and monoclinic phases in Fig. 8 following the transition temperatures reported by Khan et al.[12] In the cubic phase, $g_1, g_2$, and $g_3$ deviate significantly from the spin-only value 2.003,[35] a direct result of the strong spin-orbit interaction in $Ir^{4+}$ ions.[36] This SOC entangles spin and orbital degrees of freedom, shaping the electronic and magnetic properties of the system.[37,33] In the cubic phase, $g_1$, $g_2$ and $g_3$ show only slight temperature dependence. On entering the tetragonal phase, their temperature dependence changes noticeably. However, the change across tetragonal-monoclinic transition is subtle. In contrast, the hyperfine coupled $g_4$ shows a clear shift upon entering the monoclinic phase. Further, we comment on the hyperfine coupled $g_4$ mode. This mode is absent in the high-symmetry cubic phase but emerges deep within the tetragonal phase and persists upon further cooling through the tetragonal-to-monoclinic transition. The absence of $g_4$ in the cubic phase suggests that it is symmetry-forbidden in environments with high structural symmetry.

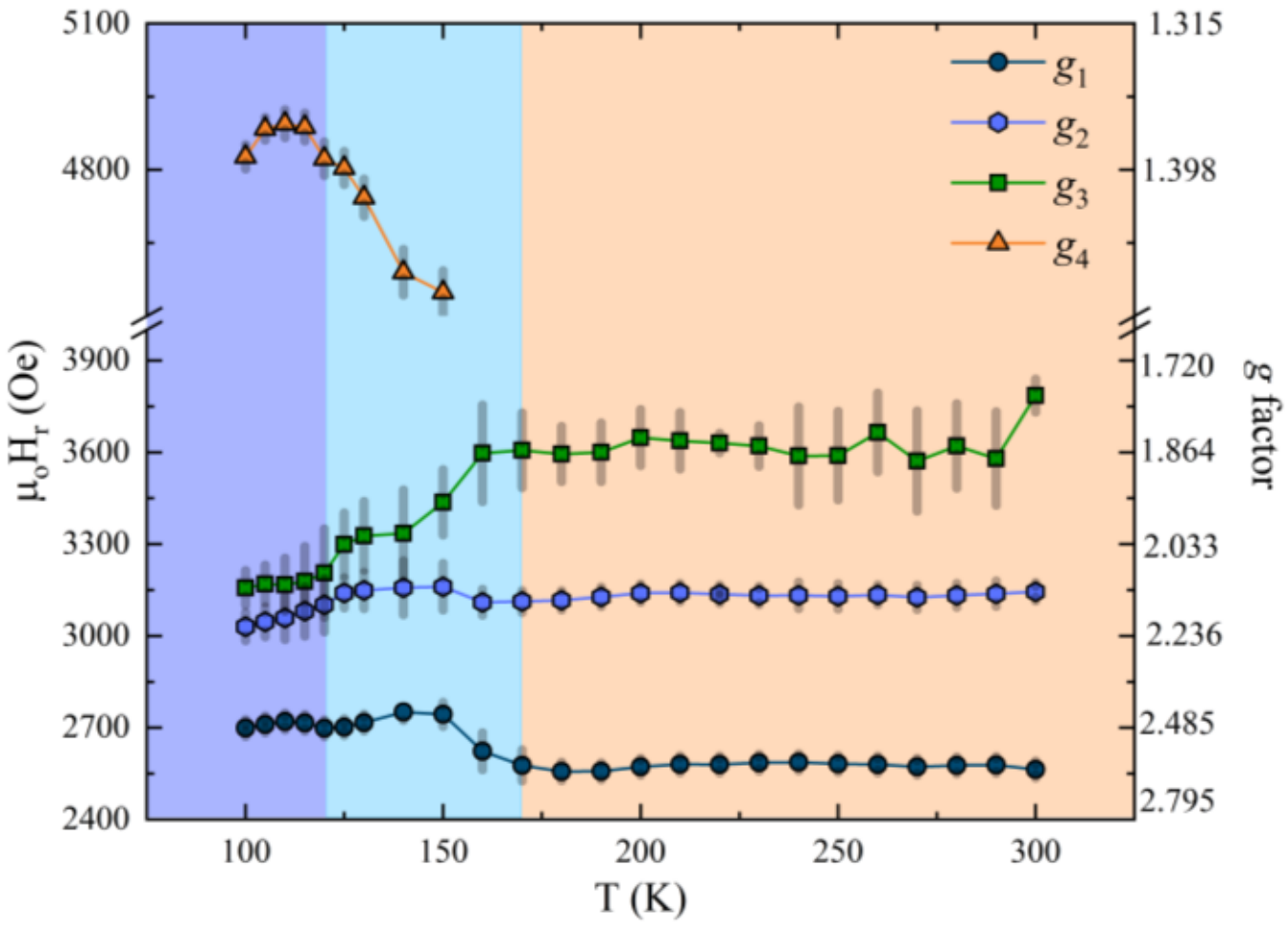

**Fig. 8** Temperature dependence of the resonance fields ($\mu_0 H_r$) and the corresponding $g$-factors for the four EPR modes. The fourth EPR mode ($g_4$) is negligible above 150 K. The shaded colors represent the different phases. From high to low temperature, they correspond to cubic, tetragonal, and monoclinic structures. The transition boundaries are drawn from the ref.[12] The error bars were obtained from the standard nonlinear least-squares fitting.

We now turn to the linewidth of the modes, focusing on the two most intense modes $g_1$ and $g_2$. Their linewidths are nearly temperature-independent above ~170 K in the cubic phase but sharply decreases with reducing temperature below 170 K, as shown in Fig. 9(a). The observed EPR linewidth behavior cannot be explained by the traditional Orbach process, which involves spin relaxation through phonons.[38] In such a case, the linewidth is expected to increase with temperature due to enhanced spin-lattice relaxation. However, in the cubic phase, the linewidth remains nearly temperature-independent, ruling out this mechanism. A sudden narrowing of the linewidth is observed below 170 K, coinciding with the cubic-to-tetragonal structural transition. This behavior likely reflects the onset of antiferromagnetic (AF) fluctuations, which average out local magnetic field inhomogeneities and lead to exchange narrowing. This indicates the presence of AF correlations well above the Néel temperature $T_N$. The EPR results are consistent with Raman measurements, where the $E_g$ mode shows anomalous linewidth broadening attributed to spin-phonon interactions in the presence of AF fluctuations. They also align with the observed phonon softening and anomalous lattice anharmonicity in the phonon energy versus temperature behavior.

We now discuss an intriguing result related to the symmetry of the $g_3$ mode's spectral weight. From the fitting, the $g_1$ and $g_2$ modes are found to be symmetric. However, as shown in Fig. 9(b), the symmetric component of the $g_3$ mode sharply drops to a negligible value upon entering the tetragonal phase, while the antisymmetric component emerges and evolves continuously with decreasing temperature, down to the lowest recorded temperature of 100 K. This temperature-dependent evolution of the $g_3$ mode below the cubic-to-tetragonal phase transition highlights the significant role of spin-lattice coupling in these phases. The origin of the antisymmetric component in the tetragonal and monoclinic phases appears to be intrinsic, as it emerges below the cubic phase transition and cannot be attributed to experimental artifacts, such as instrumental dispersion effects. A plausible explanation is that pronounced magnetic anisotropy, enhanced by strong SOC in $Ir^{4+}$ ions, breaks the symmetry of spin relaxation pathways, resulting in an antisymmetric EPR line shape.

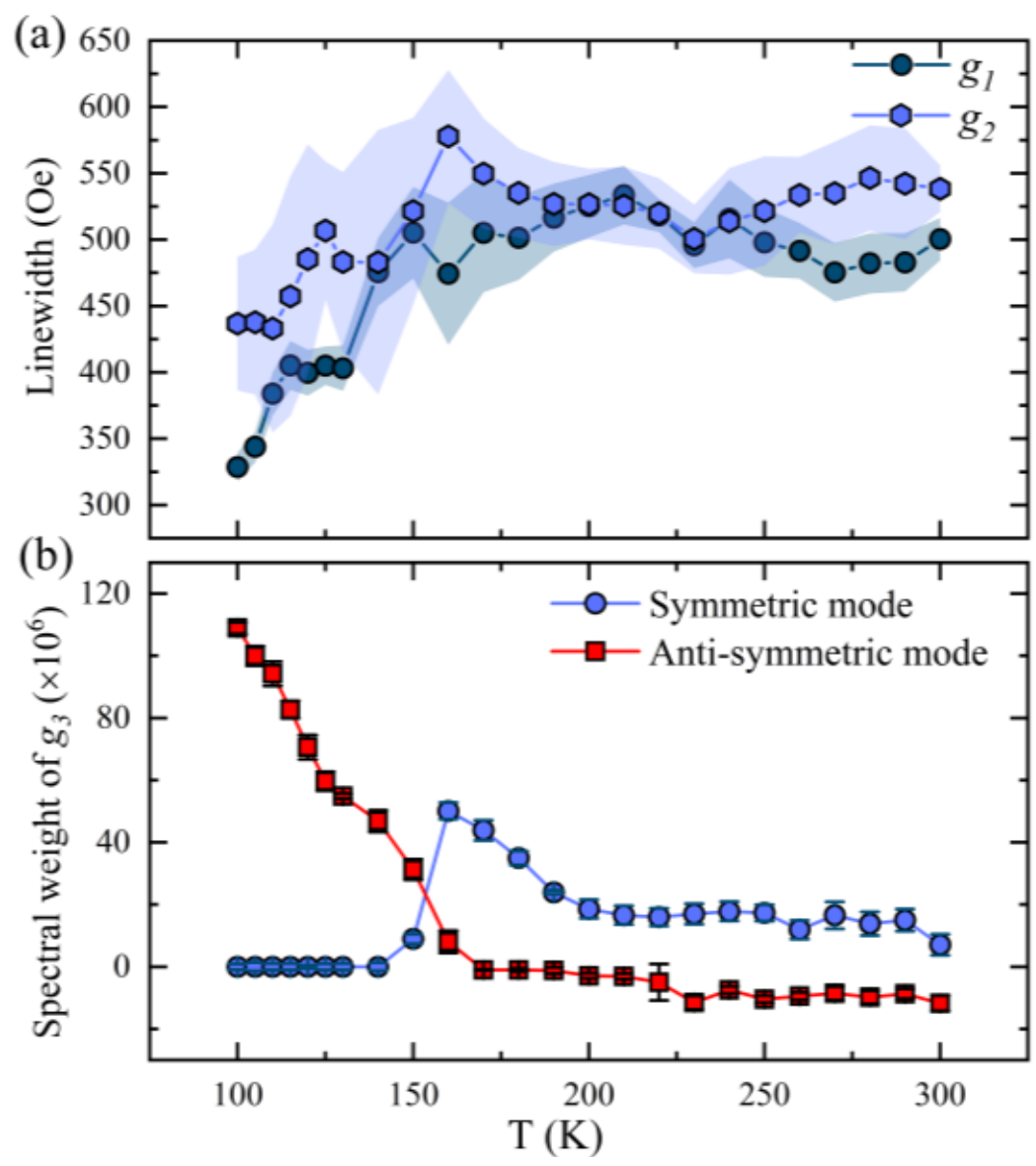

**Fig. 9** (a) Linewidth of two most prominent modes ($g_1$ and $g_2$) of KIB as a function of temperature. (b) Spectral weight of symmetric and anti-symmetric lorentzian profile of $g_3$ mode. The error bars were obtained from the standard nonlinear least-squares fitting.

## Conclusions

$K_2IrBr_6$ serves as a platform to investigate how strongly lattice vibrational properties and local magnetization ($j_{\mathrm{eff}} = 1/2$) are sensitive to the underlying lattice symmetry in the context of its structural transitions, low-temperature magnetism, and strong spin-orbit coupling. It undergoes successive structural phase transitions from a high-symmetry cubic phase to a tetragonal phase below 170 K, and then to a monoclinic phase below 122 K, driven by cooperative deformations, tilts, and rotations of the $IrBr_6$ octahedra.[12] These structural changes accompany a transition from a high-temperature paramagnetic state to an antiferromagnetic phase below $T_N \approx 16$ K.

Raman spectroscopy reveals that lattice dynamics are significantly influenced by evolving magnetic correlations. While the observed phonon modes align well with DFPT calculations, their linewidths and energies exhibit pronounced deviations from conventional anharmonic behavior. Anomalous linewidth broadening begins far above $T_N$, which indicates that dynamic spin-phonon interactions are activated by strong antiferromagnetic (AF) fluctuations. This interaction enhances lattice anharmonicity across structural transitions, emphasizing a strong coupling between phonons and fluctuating spin degrees of freedom in this halide system.

Complementary insights are obtained from EPR spectroscopy, which probes the static magnetic environment of $j_{\mathrm{eff}} = 1/2$ spins. EPR results reveal an anisotropic $g$-factor and a narrowing of linewidth, along with shifts in resonance fields, upon entering the low-symmetry tetragonal and monoclinic phases. These changes signal the emergence of persistent spin-lattice correlations, where reduced lattice symmetry enhances exchange interactions and magnetic anisotropy. The alignment between structural distortions and the evolution of the spin environment highlights a robust spin-lattice coupling, distinct from the dynamic spin-phonon processes observed via Raman.

Together, these spectroscopic findings illustrate that in $K_2IrBr_6$, spins are not free but remain intricately linked to the lattice through spin-orbit entanglement. This entanglement mediates both dynamic (phonon-coupled) and static (lattice-coupled) interactions, making $K_2IrBr_6$ an exemplary system to explore coupled spin-lattice phenomena in spin-orbit coupled halides. These results underline how the choice of ligands and the ease of structural deformations in halide frameworks can be used to control spin-lattice interactions, offering a strategy to design quantum materials with tunable magnetic properties.

Furthermore, related halide compounds such as $K_2IrCl_6$ have been proposed to host nodal-line spin-liquid phases, where spin-orbit coupling and exchange frustration stabilize extended spin correlations without long-range magnetic order. While $K_2IrBr_6$ has not yet been explored in this context, the persistence of spin correlations and lattice anomalies observed here suggests it may exhibit similar correlated spin dynamics, warranting further investigation.

## Author contributions

S.B. acquired and analyzed XRD, Raman, and EPR data. A.A., M.Z. and B.K.M. performed lattice dynamics calculations. S.B. and S.K. did magnetization measurements and the corresponding data analysis. R.K. and S.F. developed the low-temperature Raman scattering setup. J.K.A. and A.G. performed initial Raman measurements for characterization using a commercial Raman spectrometer. K.S. conceived the idea and supervised the project. All authors contributed to the writing of the manuscript.

## Conflicts of interest

There are no conflicts to declare.

## Data availability

The data supporting this article are available within this article and have been included as part of the ESI. †

## Acknowledgements

This work is supported by a Core Research Grant from SERB, a YSRP grant from BRNS and an INSPIRE Faculty Award from DST in India. K.S. further acknowledge the seed funding from IIT Delhi. We also acknowledge the FIST Grant (SR/FST/ETI/2020/628). S.B. acknowledges her PhD fellowship from the CSIR India. All authors acknowledge the support from NRF, CRF, HPC facilities at IIT Delhi and the Institute Instrumentation Centre at IIT Roorkee. K.S. thanks N. Khan for multiple scientific discussions on $K_2IrBr_6$.

## References

1 W. Witczak-Krempa, G. Chen, Y. B. Kim and L. Balents, *Annu. Rev. Condens. Matter Phys.*, 2014, **5**, 57–82.

2 G. Jackeli and G. Khaliullin, *Phys. Rev. Lett.*, 2009, **102**, 017205.

3 Y. Noda, T. Ishii, M. Mori and Y. Yamada, *J. Phys. Soc. Japan*, 1980, **48**, 1279–1287.

4 R. L. Armstrong, *Phys. Rep.*, 1980, **57**, 343–396.

5 H. W. Willemsen, C. A. Martin, P. P. M. Meincke and R. L. Armstrong, *Phys. Rev. B*, 1977, **16**, 2283.

6 L. Savary and L. Balents, *Reports Prog. Phys.*, 2016, **80**, 016502.

7 B. J. Kim, H. Jin, S. J. Moon, J. Y. Kim, B. G. Park, C. S. Leem, J. Yu, T. W. Noh, C. Kim, S. J. Oh, J. H. Park, V. Durairaj, G. Cao and E. Rotenberg, *Phys. Rev. Lett.*, 2008, **101**, 076402.

8 Y. Singh and P. Gegenwart, *Phys. Rev. B*, 2010, **82**, 064412.

9 T. Dey, A. Maljuk, D. V. Efremov, O. Kataeva, S. Gass, C. G. F. Blum, F. Steckel, D. Gruner, T. Ritschel, A. U. B. Wolter, J. Geck, C. Hess, K. Koepernik, J. Van Den Brink, S. Wurmehl and B. Büchner, *Phys. Rev. B*, 2016, **93**, 014434.

10 N. Khan, D. Prishchenko, Y. Skourski, V. G. Mazurenko and A. A. Tsirlin, *Phys. Rev. B*, 2019, **99**, 144425.

11 Q. Wang, A. de la Torre, J. A. Rodriguez-Rivera, A. A. Podlesnyak, W. Tian, A. A. Aczel, M. Matsuda, P. J. Ryan, J.-W. Kim, J. G. Rau and K. W. Plumb, *Phys. Rev. X*, 2025, **15**, 021021.

12 N. Khan, D. Prishchenko, M. H. Upton, V. G. Mazurenko and A. A. Tsirlin, *Phys. Rev. B*, 2021, **103**, 125158.

13 G. Kresse and J. Furthmiiller, *Comput. Mater. Sci.*, 1996, **6**, 15–

50.
14 G. Kresse and J. Furthmu, *Phys. Rev. B*, 1996, **54**, 11169.
15 P. E. Blöchl, *Phys. Rev. B*, 1994, **50**, 17953.
16 J. P. Perdew, K. Burke and M. Ernzerhof, *Phys. Rev. Lett.*, 1996, **77**, 3865.
17 S. L. Dudarev, G. A. Botton, S. Y. Savrasov, C. J. Humphreys and A. P. Sutton, *Phys. Rev. B*, 1998, **57**, 1505–1509.
18 A. Togo and I. Tanaka, *Scr. Mater.*, 2015, **108**, 1–5.
19 M. Majumder, F. Freund, T. Dey, M. Prinz-Zwick, N. Büttgen, Y. Skourski, A. Jesche, A. A. Tsirlin and P. Gegenwart, *Phys. Rev. Mater.*, 2019, **3**, 074408.
20 S. Lee, B. H. Kim, M. J. Seong and K. Y. Choi, *Phys. Rev. B*, 2022, **105**, 184433.
21 P. Stein, T. C. Koethe, L. Bohatý, P. Becker, M. Grüninger and P. H. M. Van Loosdrecht, *Phys. Rev. B*, 2023, **107**, 214301.
22 M. S. Senn, D. A. Keen, T. C. A. Lucas, J. A. Hriljac and A. L. Goodwin, *Phys. Rev. Lett.*, 2016, **116**, 207602.
23 P. G. Klemens, *Phys. Rev. B*, 1966, **148**, 845.
24 J. Menéndez and M. Cardona, *Phys. Rev. B*, 1984, **29**, 2051–2059.
25 K. Sen, Y. Yao, R. Heid, A. Omoumi, F. Hardy, K. Willa, M. Merz, A. A. Haghighirad and M. Le Tacon, *Phys. Rev. B*, 2019, **100**, 104301.
26 O. Abdul-Aziz, D. Wolverson, C. J. Sayers, E. Carpene, F. Parmigiani, H. Hedayat and P. H. M. van Loosdrecht, *J. Mater. Chem. C*, 2024, **12**, 11402–11411.
27 T. T. Thu Nguyen, Y. Kim, S. Bae, M. Bari, H. R. Jung, W. Jo, Y. H. Kim, Z. G. Ye and S. Yoon, *J. Phys. Chem. Lett.*, 2020, **11**, 3773–3781.
28 J. Khatua, S. Bhattacharya, A. M. Strydom, A. Zorko, J. S. Lord, A. Ozarowski, E. Kermarrec and P. Khuntia, *Phys. Rev. B*, 2024, **109**, 024427.
29 Q. Y. Xie, F. Xiao, K. W. Zhang and B. T. Wang, *Phys. Rev. B*, 2024, **110**, 045203.
30 R. Kumawat, S. Farswan, S. Kaur, S. Bhatia and K. Sen, *J. Phys. Condens. Matter*, 2024, **36**, 493001.
31 J. H. E. Griffiths and J. Owen, *Proc. R. Soc. London. Ser. A. Math. Phys. Sci.*, 1954, **226**, 96–111.
32 J. Khatua, Q. P. Ding, M. S. R. Rao, K. Y. Choi, A. Zorko, Y. Furukawa and P. Khuntia, *Phys. Rev. B*, 2023, **108**, 054442.
33 K. Chakarawet, M. Atanasov, J. E. Ellis, W. W. Lukens, V. G. Young, R. Chatterjee, F. Neese and J. R. Long, *Inorg. Chem.*, 2021, **60**, 18553–18560.
34 B. J. Hales, *Encycl. Inorg. Chem.*, 2008, 1–16.
35 H. Han, L. Zhang, H. Liu, L. Ling, W. Tong, Y. Zou, M. Ge, J. Fan, C. Zhang, L. Pi and Y. Zhang, *Philos. Mag.*, 2015, **95**, 3014–3022.
36 M. Ju, J. Kim and J. Shin, *Bull. Korean Chem. Soc.*, 2024, **45**, 835–862.
37 P. Warzanowski, M. Magnaterra, C. J. Sahle, M. M. Sala, P. Becker, L. Bohatý, I. Císařová, G. Monaco, T. Lorenz, P. H. M. van Loosdrecht, J. van den Brink and M. Grüninger, *Phys. Rev. B*, 2024, **110**, 195120.
38 A. Najev, S. Hameed, A. Alfonsov, J. Joe, V. Kataev, M. Greven, M. PoŽek and D. Pelc, *Phys. Rev. B*, 2024, **109**, 174406.